# Intelligent Document Processing - Methods and Tools in the real world

**Graham A. Cutting**
Independent Researcher, F
grahamcutting@cantab.net
**Anne-Françoise Cutting-Decelle**
Université de Genève/CUI, CH
anne-francoise.cutting-decelle@unige.ch

## Abstract

The originality of this publication is to look at the subject of IDP (Intelligent Document Processing) from the perspective of an end-user and industrialist and not that of a Computer Science researcher. This domain is one part of the challenge of information digitalisation that constitutes the Industrial Revolution of the 21st century (Industry 4.0) and this paper looks specifically at the difficult areas of classifying, extracting information and subsequent integration into business processes with respect to forms and invoices.

Since the focus is on practical implementation a brief review is carried out of the market in commercial tools for OCR, document classification and data extraction in so far as this is publicly available together with pricing (if known). The intuition is that, in Europe at least, the trend seems to be towards the prevalence of digitally-born documents and therefore the known problems with OCR of documents from uncontrolled sources are becoming less important.

Brief definitions of the main terms encountered in Computer Science publications and commercial prospectuses are provided in order to de-mystify the language for the layman.

A small number of practical tests are carried out on a few real documents in order to illustrate the capabilities of tools that are commonly available at a reasonable price. The unsolved (so far) issue of tables contained in invoices is raised. The case of a typical large industrial company is evoked where the requirement is to extract 100% of the information with 100% reliability in order to integrate into the back-end Enterprise Resource Planning system.

Finally a brief description is given of the state-of-the-art research by the huge corporations who are pushing the boundaries of deep learning techniques further and further with massive computing and financial power – progress that will undoubtedly trickle down into the real world at some later date. The paper finishes by asking the question whether the objectives and timing of the commercial world and the progress of Computer Science are fully aligned.

## 1. Introduction

The subject of Intelligent Document Processing (abbreviated to IDP) is very topical since it brings together many of the concerns and opportunities of today's society – digitalisation (or indeed digitisation), automation, artificial intelligence, cost competitiveness and ecology (reduction of paper). There are accordingly an enormous quantity of publications, articles, posts and blogs being issued in this respect more or less every day and the number of researchers engaged in expanding the knowledge and boundaries has exploded over the last 2 or 3 years. Most scientific publications take a "deep dive" into one specific aspect of this domain and investigate areas that are becoming less and less accessible to the general public.

This contribution takes a slightly different approach – one of the authors worked for many years in industry (financial management) but has no formal IT training or programming/development experience. The perspective is therefore in part that of a potential end-user of the technology being developed and that of an "outsider looking in". This paper will attempt to:

- Cut through the jargon and offer a layman's understanding of the main terms and acronyms that can be found;
- Describe the panoply of possible tools that can be used in IDP, ranging from the simplest (directly accessible to end-users) to the most complex "state-of-the-art" research (at the time of writing). The bibliography will provide some more in-depth specialist insights into each of the methods mentioned;

- Consider not only the open-source tools that are very often used in pure research projects but also some of the main commercial software that is represented or recommended in real-world solutions;
- Provide some views on applying these tools to real life cases and documents – from the client's perspective and not from a computer science perspective;
- Take as test examples documents in another language other than English (specifically French) – the vast majority of scientific publications taking English-language as their test/training samples and application of the tools to other languages and cultures bring different challenges;
- Offer some brief comments on leading-edge tools developed in the last 2-3 years.

## 2. Some basic definitions

### 2.1 Intelligent Document Processing

The definition of this term indicates a number of different approaches and perspectives:
"Intelligent Document Processing (IDP), sometimes referred to as intelligent capture, is a set of technologies that can be used to understand and turn unstructured and semi-structured data into a structured format.
Unlike optical character recognition (OCR), IDP uses artificial intelligence (AI) technologies such as machine learning (ML) and natural language processing (NLP) to capture, classify, and extract the most difficult-to-automate data. Robotic Process Automation (RPA) technology can then be applied to the extracted data for enhanced validation and to automatically enter it into existing applications." [1]
or:
"Any software product that captures data from documents (e.g. email, text, pdf and scanned documents) [and is then capable of working to] categorise and extract relevant data for further processing using AI technologies such as computer vision, OCR, Natural Language Processing (NLP) and machine/deep learning." [2]
Whilst it is true that "Artificial Intelligence" (AI) is undoubtedly required for difficult cases of data extraction and structuring and high-volume or comprehensive processing this paper will argue that a more trivial approach (templates or rules-based programs) can be useful in some specific circumstances. Data capture technology in any case will lie at the heart of the core competences of any IDP company and the borderline between program-driven rules and "machine-learning" may not be obvious to the outsider. Indeed the precise blend of technologies applied to solve IDP problems by commercial companies will probably be hidden behind the marketing language. This is clearly explained in "Intelligent Document Processing Buyer's Guide" from BIS-Grooper. [3]
Many vendors explain that their solution has a layer of fixed rules and templates before some form of AI technology kicks in to solve new or difficult cases.
This paper will therefore adopt the convention that "Intelligent" in IDP simply means a machine/computer being substituted for a human being in certain tasks – whether the machine is trying to imitate the behavior of humans or not. People carrying out repetitive tasks do not always have scope for initiatives, judgment or new learning!!
IDP is sometimes also called CDA (Cognitive Document Automation).

### 2.2 Some key terms

Before moving on to the more technical and abstruse terms that will be encountered it can be instructive to provide some short definitions of the main terms that crop up in the literature and publicity material :

### 2.2.1 Document

The world of AI is investigating the recognition and use of many types of electronic files – images, text, forms…
This paper will limit its scope to content-rich "documents" used for business purposes. As such it will not concern itself with images, car number-plates or other challenging AI fields.
One of the largest publicly accessible databases of heterogeneous documents that can be used for AI training is the IIT-CDIP – Illinois Institute of Technology Complex Document Information Processing Test Collection (see AI training section of this paper for more information).

This database has 16 categories of labelled documents: letter, memo, email, filefolder, form, handwritten, invoice, advertisement, budget, news article, presentation, scientific publication, questionnaire, résumé, scientific report, specification.

This paper will look more closely at the categories "invoice" and "form" given one of the authors' background in industry. Each of the 16 categories presents different specific challenges.

## 2.2.2 Document Imaging

This was the first step in the journey from a paper-bound world to a paperless one where all information is accessible, understandable and usable. In other words analogue to digital.

It involves converting paper documents/files or microfilm/fiche to digital images. Technology involved can include facsimile machines, document scanners, multipurpose printers, smartphone cameras… While the concept is much older the scanner as we know it was invented around 50 years ago in Kiel. [4]

The plethora of electronic documents produced by these technologies led naturally to a demand for Document Management Systems from the 1980s onwards.

As well as many vendors of such systems the International Standards Organisation (ISO/TC 46) has published a number of standards regarding the technical documentation, for example ISO 2709:2008 (ISO/TC 9846)[5] Format for Information Exchange and ISO 15836-1:2017 (ISO/TC 46/SC 4)[6] The Dublin Core metadata element set:

The standard for portable document formats (pdf) ISO 32000-2:2020 (ISO/TC 171) will be explored later in this publication.

## 2.2.3 Document Capture

The next step beyond document management was to make use of the embedded information in the document – for classification, analysis or automation purposes.

The underlying technologies began to see commercial use in the 1970s – barcodes and barcode readers are one example. For the use-cases quoted in this paper we will look more closely at OCR [7] (Optical Character recognition).

## 2.2.4 Optical Character Recognition (OCR)

This involves the electronic conversion of _images_ of typed, handwritten or printed text into machine-encoded and machine-usable text.

The process is subject to many pitfalls and errors and even in optimal conditions is rarely 100% accurate [8].

In the past OCR was mainly carried out on flat-bed scanners. More recently taking pictures with mobile phone cameras has become a convenient way of scanning physical documents.

This paper will outline some of the work that has been done to improve accuracy and some practical conclusions.

## 2.2.5 Machine-Learning (ML)

Machine learning is the study and application of computer algorithms that can improve automatically through experience and by the use of data. More formally according to Tom M. Mitchell " [9] "A computer program is said to learn from experience E with respect to some class of tasks T and performance measure P if its performance at tasks in T, as measured by P, improves with experience E."

The learning may be:

- _**Supervised**_ where a human provides advice, and in some cases the answers, with respect to deviations from the "ground truth ". A computer program that provides an algorithm with all the possible steps to solve a problem is, strictly speaking, not "learning".
- _**Unsupervised**_ where the machine learns automatically

We might want to ask the question, as did Alan Turing, whether "can machines think" should be replaced by "can machines do what we (as thinking entities) do [10]".

The result provided by the machine may be analytical (retrospective) or predictive

## 2.2.6 Deep Learning

Deep Learning is a variety of Machine Learning that uses multiple layers in different types of neural networks. This allows progressively higher levels of abstraction – for example moving from lower-levels that can recognize pixels to a higher level that can recognize a face.

## 2.2.7 Artificial Intelligence (AI)

This is the form of intelligence that can be demonstrated by machines as opposed to living beings. The field can be defined as the study of "intelligent agents" – any system that perceives its environment and takes action that maximise its chance of achieving its goal [11].
The result may be that the machine can achieve results by a process that resembles the cognitive function of humans even if this is not strictly the case. The relationship between Artificial Intelligence, Machine Learning and Deep Learning can be represented as follows in Fig. 1:

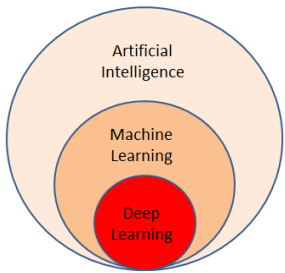

**Fig. 1 – a simple representation of the commonly reported relationship between Artificial Intelligence, Machine Learning and Deep Learning**
Some sources dispute this relationship and describe overlaps rather than subsets.

## 2.2.8 Ground Truth

This involves checking the results of machine-processing against a benchmark in the real world. This may be provided by a human or domain expert assessment of the "truth". For example an OCR processing of a printed document will be compared to the original text.

## 2.2.9 Robotic Process Automation (RPA)

IDP and RPA are complementary technologies. One can think of RPA as doing the heavy lifting and more mechanical tasks such as dealing with structured documents or mass processing of computer formats. IDP will complete the process for less well-defined or less structured documents. In practical terms this paper does not differentiate between the two since both techniques may need to be applied to the total population of documents.

## 2.2.10 "Structured", "Semi-structured" and "Unstructured" Documents

There is not universal agreement about these categories. Some practitioners and researchers consider many "semi-structured" documents as "structured" – perhaps because they are less interesting from a theoretical point-of-view or because the term varies according to the context of use-case.
 However the authors have adopted the following definitions:

- **"Structured"**
  An official (government…) form or a form from a known supplier or organisation.  The format, within a given time-frame remains fixed and predictable. A tax form, a loan application form from a specified bank or a national identity card or passport for a particular country could be examples.
- **"Semi-structured"**
  A document from a well-defined class of documents with fairly predictable types of information to be found but considerable variations from one issuing organization to another. The classic example is invoices (unless these have been standardised/structured by the government or the customer (e.g. EDIFACT (https://unece.org/trade/uncefact/introducing-unedifact) , Faktur-X – see section 8.1.1).
- **"Unstructured"**

Documents with text, images, symbols and no pre-defined format or structure. Some examples: scientific publications, legal contracts, e-mail, Wikipedia…

## 2.2.11 Different types of "Portable Document Format" documents

Portable Document Format or PDF is a format developed by Adobe in 1992 to present documents in a manner independent from the original software or operating system. After becoming a de facto standard in the business world the format(s) were standardised in 2008 (ISO TC171). The latest version is ISO 32000-2:2020 [12].

There any many features included in the different versions of this standard but the main families or types of PDF are:

- **"Machine-generated PDF" or "real PDF" or "native PDF" or "PDF/A"**

PDF documents that have been created digitally by printing to PDF from another software application and have an underlying or embedded text layer of elements that can be directly extracted from the document. In the PDF/A version the requirements include colour management guidelines and support for embedded fonts. It is not useful, and indeed may be counter-productive to scan these "semi-intelligent" documents.

- **"PDF Image"**

Can be the result of a scan of a paper document or a computer conversion of an image into a PDF format. Text cannot be extracted without going through an OCR process.

- **Searchable PDF**

This is basically a scan of a hard-copy document (or a computer conversion of an image) that has gone through an OCR process. The resulting PDF file has 2 layers: one with the image and one with recognizable text that can be manipulated just like in "real PDF".

For practical purposes a "searchable PDF" can be aggregated with a "real PDF" since after OCR it can be handled in the same way.

PDF forms, "real" or "image" have become the standard input format into Intelligent Document Processing pipelines. One specific type of PDF/A that is becoming more prevalent is the PDF/A-3 which has a layer of xml describing the content and text embedded in the document as XML. This makes the subsequent computerised use of the information a trivial case of extraction and mapping (cf. Faktur-X described later in this paper).

## 2.2.12 Regex

Regex or "Regular Expressions" are a sequence of characters that specifies a search pattern. It can be used in lexical analysis and exists in different flavours and syntaxes of which PCRE (Perl Compatible Regular Expressions) is perhaps the most common. Many languages or data extraction tools incorporate Regex components and libraries. This forms part of the "rules-based" approaches.

A sample snippet of Regex (to find a date in US MMDDYYYY format on the same line as "Invoice Date") could be:

Invoice Date\s+((0[1-9]|1[0-2]).(0[1-9]|[12]\d|3[01]).([12]\d{3}))

## 2.2.13 Object Detection

Some ML tools treat a scanned PDF as an image. Object detection through computer vision techniques is the problem of finding and classifying a variable number of objects in an image and then drawing bounding boxes around each entity field. The machine is "trained" in this process by referring to a training set of images.

For example:

- "this is (probably) an invoice"
- "the text string at x/y, x1/x2 is (probably) a due date"

Then OCR is performed to recognize and extract the content from the bounding box.

## 2.2.14 Natural Language Processing (NLP)

Some ML tools adopt a different approach and perform a semantic analysis of the text on a PDF page (after carrying out OCR if necessary). State-of-the –art NLP approaches seem to outperform Object Detection approaches according to many recent publications.

Some approaches combine Object Detection and NLP components, and indeed sometimes rules-based approaches as well.

## 3. Optimal/possible computer treatment of types/categories of documents

Putting all of this together we can summarise the most fruitful avenues for tackling IDP as follows in Table 1. Note that this paper has considered "unstructured" documents as out-of-scope given the focus on a limited number of types of business documents.

**Table 1 – main approaches to tackling computer treatment of structured and semi-structured documents**

|  | **Real PDF** | **PDF Image** |
|---|---|---|
| Structured | Fixed Template | Mapping image to Template + OCR<br>OCR + Regex<br>OCR + Object Detection + ML<br>OCR + NLP + ML |
| Semi-structured | Multiple templates (by supplier or issuing organisations)<br>Regex<br>NLP + ML | OCR + Regex<br>OCR + Object Detection + ML<br>OCR + NLP + ML |

A flow chart of the different options can be represented below in Fig 2.

## 4. The megatrends

- Thousands of scientific publications (keywords in Google Scholar).
- Hundreds of vendors from huge multinationals to small start-ups.
- Hundreds of scientific conferences.
- An explosion of interest in this field since 2018 driven partly by the expansion of the possibilities of the technology and partly by the need of industrial companies to gain in agility and resilience since the start of the Covid pandemic.

The drive to go "paperless" seems to have somewhat different patterns in Europe and in the U.S.:

- In Europe, where governments are quite heavily centralised the initiative can come top-down. For example:
    - Tax declarations and other official forms (France)
    - Passports (UK)
    - Or from standards and industry consortia – e.g. EN 16931-1 (Faktur-X) – see Section 8.1.1 for more details
- In the US where government is less centralised the driving force may be more bottom-up:
    - ROI (Return on Investment)
    - $$$$$$
    - Facilitated by technology

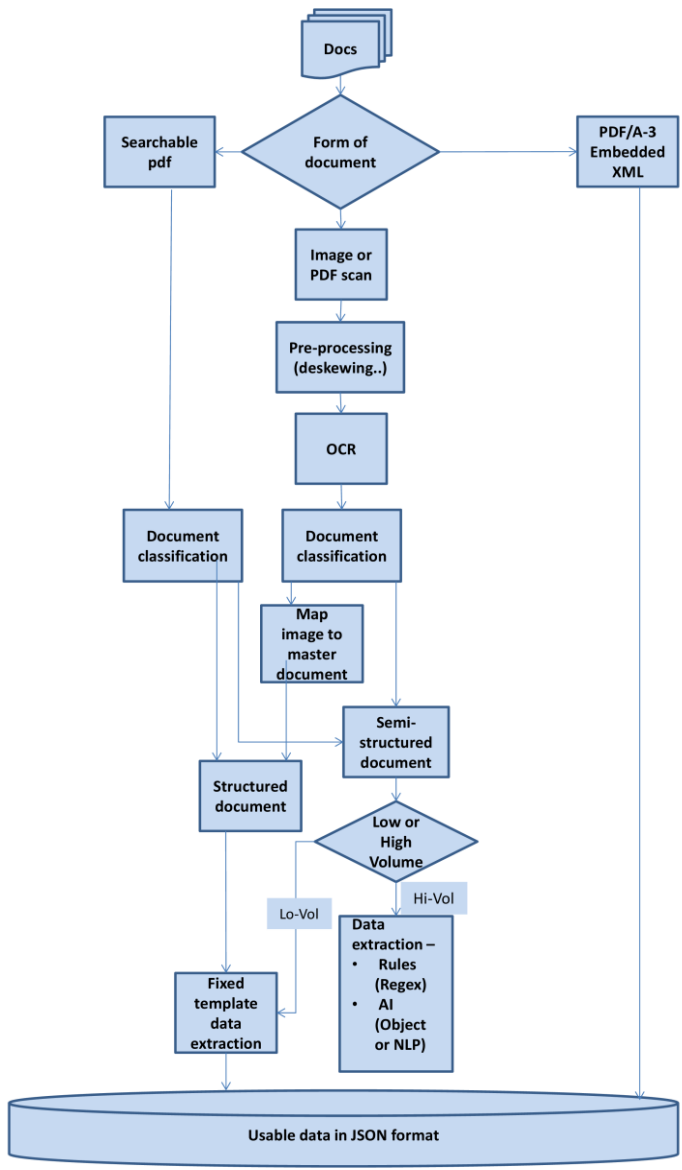

**Fig 2 A flow-chart of the different options for computer treatment of structured and semi-structured documents**

## 5. The most common use-cases

According to the web-sites of the main vendors of business document automation:

- Accounts Payable especially supplier invoices;
- US tax forms (1040, W-2, W-4…);
- Bank loan applications;
- Recognition of ID cards (driving licenses) and passports.

There are many other more niche sectors such as résumés/CVs, logistics documents….

## 6. The main types of suppliers/vendors of IDP/OCR solutions

1. Free online or offline tools
2. Open-Source tools

3. Stand-alone moderately-priced packages aimed at end-users
4. Higher-end packages with SDK capabilities
5. SaaS with or without APIs
6. ERP suppliers
7. Core technologies (GAFAM….) (Google, Apple, Facebook, Amazon, Microsoft)
8. Niche players

Brief comments will be made about the apparent capabilities of each of these categories of suppliers and more in-depth examples will be provided in the practical case-study for the categories 1-3

## 6.1 Free online and offline tools

An Internet search for online tools that can convert images or pdf files ("real" or image-based) into something more usable will rapidly reveal a long list of potential candidates for OCR or pdf conversion.
They are of very variable quality and will struggle in many use-cases.
In general:
- Plain text in standard fonts on a white background can be converted with a high degree of success (for example for onward translation into another language – sufficient at least to get the real sense of the meaning)
- Forms and tables cause more difficulties although there are some specialized tools that can perform reasonable OCR on tables if they are well demarcated
- Embedded images are also tricky

It can be noted that the very latest versions of LibreOffice Writer and Calc (and Microsoft Word/Excel) can convert very successfully searchable PDF documents (text or spreadsheets respectively) *if* they have been created originally by similar office suite applications.
However the precept that "you get what you pay for" still applies generally and if one wants more accuracy, more flexibility or onward data integration the free tools are outgunned by much commercial software or bespoke open-source programmed developments.

## 6.2 Open Source tools

Most Computer Science publications concentrate on this category of tools since they are free and also offer an open playing field for research and experimentation.
The best known tool for OCR is Tesseract (there are several alternatives) and other tools exist for extracting text from searchable PDF, for Natural Language Processing or ontological analysis.
Tesseract can be combined with OpenCV (for computer vision and Pattern Recognition), Tensor Flow/Keras  or PyTorch in machine-learning developments or research projects.

### 6.2.1 Advantages of Tesseract and associated libraries

- Vast library of tools, scripts and possible interfaces;
- Huge community of users and potential support;
- Several programming possibilities – Python, C++, Java;
- Can incorporate Machine Learning, AI…. (TensorFlow) ;
- Case-studies of very ambitious and complicated projects accomplished;
- Can be applicable to research projects with a limited scope whether from a technical perspective or just domain-specific.

### 6.2.2 Disadvantages of Tesseract and associated libraries

- Probably not the best-in-class OCR engine;
- Steep learning curve, high requirements on IT expertise and projects can rapidly become quite complicated;
- Can consume a lot of resources before delivering end-results;
- Out-of-the-box or configurable solutions may be better for commercial applications with finite resources or timeline.

A survey of the State-of-the-Art (SOTA) according to scientific publications will be made in Section 10 but very few Open-Source projects conclude with a finished product.

For obvious reasons commercial firms do not reveal all of the elements of their "technology stacks" but it seems that many of them are mixing ingredients from open-source tools, other commercial products and their own hand-crafted programs and reasoning engines.

Much of the development research in this field is openly published on the IT programmers' forums Github, Stackoverflow, Kaggle, Huggingface etc.

## 6.3 Stand-alone moderately-priced packages aimed at end-users

It can be useful to employ a dedicated commercial package if one needs immediate high-grade concrete results from OCR or PDF extraction. This can apply to:
- a private individual or a small company for a specific limited use-case;
- a research institution that wants to investigate subsequent steps in the workflow without waiting for the output from a longer-term research project.

Some software packages that have been tested (in trial or payable versions):

- Able2Extract (https://www.investintech.com/prod_a2e.htm)
- Wondershare PDFElements (https://pdf.wondershare.com/)
- ABBYY Finereader (https://pdf.abbyy.com/)
- Kofax Omnipage (https://www.kofax.com/products/omnipage)
- Adobe Acrobat Professional (https://www.adobe.com/acrobat.html)

There are other "legacy" OCR suppliers that have been bought up and integrated into more complete solutions.

The price point for this type of tool is typically €100-€200.

Adobe Acrobat professional is no longer available for purchase of a lifetime license – they now offer a subscription model of €18/month.

The concrete results of several of these solutions are shown in Section 9 of this paper.

## 6.4 Higher-end packages with SDK capabilities

If one requires integrating data extraction from a document into a workflow an Application Protocol Interface (API) will be needed which can be activated and customised through a Software Development Kit.

These are offered by the suppliers of the stand-alone packages as well as other players who focus on the complete document workflow.

The following features can be found:
- Claim AI capability and even more accurate OCR processing;
- More highly developed pre-processing tools – deskewing etc.;
- Standard or customised templates ;
- Some claim « intelligent » self-learning templates;
- Output in XML, JSON etc. ;
- Some claim automatic form extraction ;
- Many customer use-cases quoted;
- For small and medium companies often limited to a few use-cases.

Some of the solutions encountered:
- (Kofax) Omnipage Capture SDK
- ABBYY Cloud OCR SDK
- Docparser (https://docparser.com/)
- Bytescout SDK (https://bytescout.com/)

Pricing is typically $5000-$15000 or « please contact us » or Cloud version + consulting + IT resources

A project for one family of semi-structured documents (for example supplier invoices) may cost $50000- $100000 (source BIS-Grooper).

Many of the players will partner with other suppliers who are specialists for different parts of the workflow – e.g. Abbyy who partner with UiPath (https://www.uipath.com/) for a more complete RPA solution.

Bytescout is an interesting case since they offer for free a package "PDF Multitool" that can act as a "taster" or "test-bench" for the capabilities of their SDK solutions:
- OCR quality close to that of commercial end-user packages;
- Several pre-processing filters including de-skewing, contrast.;
- Output in TXT, XML, JSON;
- Accepts templates for Zonal OCR – provided that the document dimensions and positions remain fixed;
- Can also extract text/values/dates based on search criteria including Regex;
- Can also extract tables if they are demarcated;
- Can classify documents based on fixed rules, e.g. Find Term "Invoice" or "Form W4".

For most of the other suppliers one needs to sign up as a commercial prospect with a valid (profitable) use-case in order to test the capabilities.

## 6.5 Software as a service (SaaS)

Some players are exclusively SaaS (software as a service) – pay for what you use with degressive prices according to volume.
This approach can be used for stand-alone documents or APIs.
This is the main business model for the GAFAM (Google, Apple, Facebook, Amazon, Microsoft) and also Adobe PDF Extract API.
It is also favoured by some niche or specialist players since it provides a more regular income stream than a lifetime license and may offer a psychological barrier to cancellation.
Free trials are sometimes possible but it may be necessary to provide credit card details beforehand.
Pricing for OCR + data extraction is typically $0.02 - $0.06 per document.

## 6.6 ERP suppliers

Naturally several of the players in the ERP (Enterprise Resource Planning) integrated business solutions field have started to propose solutions for document extraction in order not to leave all of this market to other supplies.
SAP and Microsoft seem to be active in this field but a company would need already to be a customer of these suppliers in order to have a better idea of their capabilities.

## 6.7 Core technologies (GAFAM…)

Many of the underlying technologies and leading-edge scientific research have now been appropriated by the GAFAM companies – or at least specifically Google/Amazon/Microsoft. This reflects their immense financial resources as well as their strategic drive to obtain dominant market positions.
Probably best in class for many technologies (facial recognition, hand-writing, obscure fonts – their tools have been trained on millions of documents). They are all actively involved in AI (as are Facebook and Apple).
Their offers extend to complete business processes, data and document hosting, workflows…
Low volumes for limited use-cases are affordable for small companies
Business integration probably only feasible for large companies
Their offers are branded in different ways as sub-sets of the larger corporation – Google Cloud Vision (https://cloud.google.com/vision), Amazon Rekognition or AWS (https://aws.amazon.com/rekognition/), Microsoft Azure (https://azure.microsoft.com/en-us/).
Google Cloud Vision offers the possibility to test/make a demonstration for free of individual documents and visualize the JSON output.

## 6.8 Niche players

In the vast suppliers landscape there are many smaller companies, start-ups or niche players.

They combine many of the claimed features of the previous categories but their real capabilities are not always apparent (« please contact us »). They mostly claim AI technology.

Their pricing is usually not transparent

Their real use-cases may be limited to a few domains (they mostly always quote supplier invoices plus a few others). They may have a few tens of employees at most with current (2021) turnover not exceeding single-digit $M.

Some names that crop up often in Internet searches, comparative reviews or simply because they have a very active "blog" presence (as a commercial "taster"):

- Docsumo (https://docsumo.com/)
- Filestack (https://www.filestack.com/)
- Nanonets (https://nanonets.com/)
- Parashift (https://parashift.io/)
- Rossum (https://rossum.ai/)

This list is far from comprehensive and it is difficult to establish a solid opinion about their capabilities to deliver a working solution – but many of them do quote successful use-cases and big companies using their technology.

## 7. Domain-specific glossaries, dictionaries, thesauruses, and ontologies

Many of the studies on IDP and document extraction refer to validation of the text through dictionaries, glossaries and domain-specific ontologies. But they are rarely specific about the tools used or the measurable value-added of such methods.

A number of tools include language dictionaries or defined "cultures" in order to cover:

- Specific words found in different languages (Tesseract). These packages may also include training packages to assist in machine-learning in these languages
- The accents found in non-ASCII Latin-based languages (ABBYY Finereader)
- The date-formats, decimal points, currency symbols ($, €, £..) for different countries (Bytescout)

One of the most recent, and highly developed, tools in NLP through AI – LayoutXLM (Microsoft) as an evolution from LayoutLM – incorporates as one of its main features capabilities for multilingual, multi-culture processing.

More-restricted word lists or thesauruses can enhance text recognition by limiting the potential scope of document classification or data extraction (publications……).

Some approaches in machine-learning may seek to define by a statistical analysis or other the terms that may be retained as the target population amongst the whole "bag-of-words" (publications….).

Ontologies can also add value by defining the structure of potentially valid terms and the relations between them including a multi-dimensional aspect such as cross-validation with a user profile. The best-known Open-Source tool is probably Protégé – other free of commercial tools exist on the market.

Software such as GATE (General Architecture for Text Engineering) [13] (open-source) can be used to apply ontologies to the "corpus" (the text in question) in order to mark up, annotate or extract relevant information from the text and arrive at a sustainable information management system.

Fonduer is also used by researchers in an Open-Source Python environment to create KBCs (Knowledge Base Constructions).

This whole domain seems to be a fruitful area for further research.

The gold standard in ontologies is quoted by some authors as being MeSH [14] (Medical Subject Headings) which is used for, amongst other avenues, classifying medical publications. It consists of around 25000 terms and more than 200000 medicines in a field where approximate definitions cannot be tolerated. But there do not seem to be many other similar examples with the same rigorous approach in other technical domains

## 8. Some concrete results from easily available tools

Since the objective of this paper is not to carry out research in a narrowly-defined area of Computer Science but rather to perform a general survey of the SOTA and to assess the possible contribution of easily accessible tools a limited number of real tests have been made:

- with 6 different documents

- o 2 different tax forms (France and Switzerland)
- o 3 different invoices (2 French and one one Anglo-Saxon)
- o A page of text with "difficult" fonts

Extracts from these use cases are described below. The number of documents may not be significant from a statistical point-of-view or for machine-learning but the study can be extended in order to arrive at firmer conclusions and establish a baseline to be compared with scientific research.

The tools that have been used are;

- Able2Extract
- Wondershare PDFElements
- ABBYY Finereader
- Bytescout PDFMultitool
- Tesseract (OCR only without any "tweaks" or additional libraries
- Google Cloud Vision

## 8.1 The use cases

The complete workflow all the way through to a backend business management framework such as an ERP system.is shown below in Fig. 3

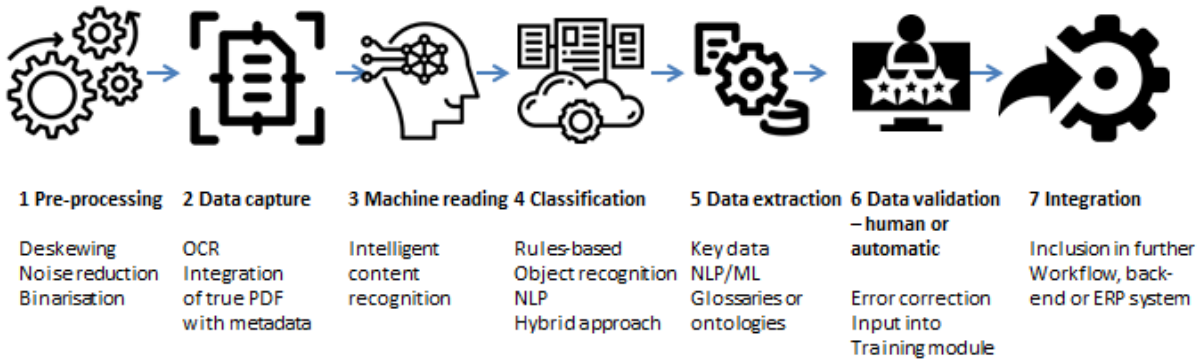

| 1 Pre-processing | 2 Data capture | 3 Machine reading | 4 Classification | 5 Data extraction | 6 Data validation – human or automatic | 7 Integration |
|---|---|---|---|---|---|---|
| Deskewing Noise reduction Binarisation | OCR Integration of true PDF with metadata | Intelligent content recognition | Rules-based Object recognition NLP Hybrid approach | Key data NLP/ML Glossaries or ontologies | Error correction Input into Training module | Inclusion in further Workflow, back-end or ERP system |

Icons made by XXXXXXX from www.flaticon.com : 1. Becris 2. Orvipixel 3. Becris 4. Eucalyp 5. Orvipixel 6. Eucalyp 7. No attribution required

**Fig. 3 The full end-to-end workflow of IDP**

The steps that can be sampled and tested with stand-alone commercial tools are 1-5. The full automation process and architecture is beyond the scope of this paper.

Real life fully-documented implemented use-cases are rather rare for evident reasons of confidentiality and also perhaps because business people are not often involved in the research project. The exceptions will be described as well as listed in the bibliography. The authors have therefore limited themselves to looking at a couple of perspectives that that know personally and illustrating the output of some readily available tools. Many vendors quote the names of big companies with successful implementations of their products but it is difficult to know in concrete terms the techniques that have been deployed.

### 8.1.1 Large industrial firms

Some vendors illustrate on the websites some real-life "customer stories" – ABBYY is one particularly rich example with 140 use-cases but these are normally 2-3 page summaries – enough to get a flavour of the use-case but not enough to betray any "trade secrets" about how it is done.

This paper will look in more detail at supplier invoices and the "Holy-Grail" of the larger "Purchase-to-Pay" process including the triple matching of invoice, purchase order and goods received note.

One of the authors' personal experience of a large industrial manufacturing company is that the classic paper-OCR-classification-extraction cycle may no longer be totally relevant for such enterprises for a number of reasons:

- Since the early 1990s such companies have been pushing to use electronic document communication means such as EDIFACT [15]. Even if this electronic form is rather "old school" many companies will have systems invested in this format.
- Large industrial manufacturing companies will certainly be using some type of ERP system (SAP, Oracle…). These systems are very heavy to install and maintain but once configured provide a high degree of automation and cross-process benefits. For example suppliers will be maintained in the relevant section of the Master Data records (names, addresses, terms of payment….). As such the task of additionally formulating a supplier-specific template is not too onerous
- Such firms will also seek to deal with a relatively limited number of suppliers as "partners" (200-500 might be a typical number) for long-term win-win relationships. The process will often involve Quality Assurance procedures and Audits. The manufacturing company will often be larger than its suppliers to will have a certain "weight" in imposing agreed administrative procedures as well – for example supplier portals and communication methods. They will undoubtedly insist on Purchase Orders for all mainstream goods – which will facilitate error-free identification of line items.

So one can suspect that electronic means of communication will be prioritised and that otherwise templates by supplier are viable options. The relevant document class will be "supplier invoice" and the relevant sub-class "invoice-supplier xxx".

The major ERP vendors certainly have offers in the domain of "Document Information Extraction" example : [16] but one would need to be an actual or prospective customer of these firms to have a clearer idea of the content.
On a more general level, in Europe at least, governments, standardisation bodies and industry consortia are moving in the direction of mandatory electronic formats. The previously mentioned EN16931-1 (Faktur-X) [17] using the PDF/A-3 will be become mandatory in 2023 for large companies supplying the French government and one can expect this format to gradually replaces older formats such as EDIFACT.
Below is a snippet from the Faktur-X xlm validation file (which contains 12000 lines!):

```
<xsl:choose>
    <xsl:when    test="(ram:ID)    or    (ram:GlobalID)    or    (ram:SpecifiedLegalOrganization/ram:ID)    or
(ram:SpecifiedTaxRegistration/ram:ID[@schemeID='VA'])" />
    <xsl:otherwise>
     <xsl:failed-assert    test="(ram:ID)    or    (ram:GlobalID)    or    (ram:SpecifiedLegalOrganization/ram:ID)    or
(ram:SpecifiedTaxRegistration/ram:ID[@schemeID='VA'])">
      <xsl:attribute name="id">BR-CO-26</xsl:attribute>
      <xsl:attribute name="flag">fatal</xsl:attribute>
      <xsl:attribute name="location">
       <xsl:apply-templates mode="schematron-select-full-path" select="." />
      </xsl:attribute>
      <svrl:text>[BR-CO-26]-In order for the buyer to automatically identify a supplier, the Seller identifier (BT-29),
the Seller legal registration identifier (BT-30) and/or the Seller VAT identifier (BT-31) shall be present.</svrl:text>
     </svrl:failed-assert>
    </xsl:otherwise>
  </xsl:choose>
  <xsl:apply-templates mode="M10" select="*" />
 </xsl:template>
```

The full package comes with sample invoice formats, business rules etc.
An extract from the sample invoice (in French) indicates that identification of the key fields becomes child-play – see the recommended structure below in Fig. 4.

Fig. 4 – the header of the recommended invoice format in EN16931-1

A heavy process but ultimately effective it has been once implemented…………
In the same way online applications and data entry is becoming prevalent (tax forms etc.) which will in turn standardise the pdf or other output formats.

A more market-driven approach from a large company is described in the publication from American Family Insurers 2021 [18]. This use-case, described in detail, explains that their industry (insurance claims) encounters hundreds of thousands of paper documents annually in myriad forms. The publication explains not only their approach but also describes the final IT architecture. The volume of documents involved was sufficient for them to benefit from a machine-learning approach. Interestingly they also describe how to deal with the inevitable errors and anomalies from the OCR/classification/extraction process – involve the business end-users themselves who then re-inject the corrected results into the machine-learning process
.

### 8.1.2 Private individuals and small companies

We can now turn to the other end of the spectrum where individuals or small companies may want to go paperless but do not have the heavy requirements of an ERP system.
A private individual may only see 10-20 documents per month (invoices, medical forms, tax declarations…) and it may be sufficient for his/her needs to:
- Classify the documents (in electronic or scanned form) in a structured way so that they can be retrieved when needed
- Receive reminders for due dates (payment dates, subscription renewals, tax deadlines…)
As such these can be catered for by simpler means than "Intelligent Document Processing", for example:
- Microsoft OneNote [19] with reminder dates triggered in Outlook
- Microsoft Sharepoint [20]
- A system built around Dropbox or Google Drive…
However in the middle ground there are many middle-sized companies or entrepreneurs who could benefit from IDP and may even need to integrate invoice details into a stock management system.
The problems of this small or middle market are not new. They have already been raised by Schuster et al. 2014 [21] with "Intellix" – End-User Trained Information Extraction for Document Archiving" for SOHO (small office and home office) users. This technology was subsequently integrated into DocuWare as part of a Document Management System. The reported price-point of a minimum of $300 per month for 4 users for this system seems to be skewed in relation to alternative solutions with cheaper subscription models for small businesses.
So it can be instructive to examine what are the practical limitations of OCR and readily available tools before entrusting one's fate into the hands of vendors who may be able to deploy more sophisticated (and costly) tools as well as Artificial Intelligence. We will briefly look at, with appropriate examples, the problems of OCR in general, invoices and data extraction.

### 8.2 OCR

Complete publications, books or even conferences can be dedicated to the issue which still has some challenges – particularly "in the wild". State-of-the-art OCR systems have now largely solved the issues for professionally-scanned business documents that have been generated by machines and printed on a well-contrasted background. Handwriting can still also pose some challenges but this is beyond the scope of this paper.

The first example covers two snippets of text from a club cycling magazine article from the 1970s and no doubt type-set by a local printer. The pages have been scanned with a professional-grade scanner with automatic de-skewing so some of the possible pre-processing steps can be avoided.

The title of the article:

And the last 3 sentences of the first page:

The issues here are the unusual fonts, the language (French) and in one case the vocabulary – see the results in Table 2 below

The pieces of software that have been tested on this text (in default setting and with French selected as a possible language) are:

- Able2Extract
- Wondershare PDFElements
- ABBYY Finereader
- Bytescout PDF MultiTool
- Google Cloud Vision
- ABBYY Finereader (FR & Bluebeard font)

**Table 2 – the results of OCR performed on the title**

| Software | Result (Title) |
|---|---|
| Able2Extract | **Balade  dans le HOGGA.R ..** |
| Wondershare PDF Elements | iBalad  e   dans le H |
| ABBYY Finereader (EN) | ᴿalade dans le |
| ABBYY Finereader (FR & Bluebeard font) | Balade dans le HOGGAR . |
| ABBYY Finereader (FR,Bluebeard font & supervised learning) | Balade dans le HOGGAR . |
| Bytescout (FR) (txt file) | iBalade dans le HOGEAR .. |
| Google Cloud Vision (JSON) | "text": "Balade dans le HOGGAR ..\n" |

Conclusion – there are issues here with the font, the language setting and also confusion between 2 words that are both in the dictionary (« balade » and « salade »). The correct results are obtained by ABBYY Finereader (after some onerous manual training – one can hope that the cloud and SDK versions are more user-friendly) and Google Cloud Vision.

**Table 3 – the results of OCR performed on the final sentences**

| Software | Result (Text) |
|---|---|
| Able2Extract | ei n Mais je recrave. Dons ma précipitation, je viens de me foire avoir comme un 06butant : une épine de2 cm est fiche dans Le pneu. IL fait nuitLorsque j'arrive là un endroit de bivouac, Les gueLtos d'ImLacuLcouêne.D8sogréobLe surprise: je ne suis pas seui. |
| Wondershare PDF Elements | .. . Mats je recrève, Dons ma précipitation, je viens de me foire avoir comme un débutant : une épine de 2 cm est fichée dons LE pneu. Lorsque j'Qrrlve ô un endroit de bivouac, Les guettas d* ImLQOUlaouène. Désagréable surprise : je ne suis pas seul |
| ABBYY Finereader (EN) | ... Wals je recreve. Dans mo precipitation, je viens de me falre avoir comme un debutant : une eplne de 2 cm est flchce dons le preu. Il fait nult lorsque j'arrlve a un endrolt de bivouac, Les gueltas d'ImLaouLaouene. DesogrcabLe surprise : je ne suls pas seul* |
| ABBYY Finereader (FR) | ... Vais je recrève. Dans ma précipitation, je viens de me faire avoir comme un débutant : une épine de 2 cm est fichée dons le pneu. Il fait nuit lorsque j'arrive à un endroit de bivouac, Les gueltos d'ImLaoulaouène. Désagréable surprise : je ne suis pas seul. |
| Bytescout (FR) (txt file) (various filters such as contrast dilate and greyscale applied) | ...... Mais je recrève. Dans ma précipitation, je vlens de me faire avoir comme un débutant : une épine de % cm est fichée dons Le pneu. ∣L fait nuit Lorsque j'arrive à un endroit de bivouac, Les queltos d'ImLooulacuène. DésagréabLle surprise : je ne suis pas seul. |
| Google Cloud Vision (JSON) | `"text": "... Mais je recrève. Dans ma précipitation, je viens de me faire avoir\ncomme un débutant : une épine de 2 cm est fichée dans le pneu. Il fait nuit\nLorsque j'arrive à un endroit de bivouac, les gueltas d'imLaoulaouene.\nDésagréable surprise : je ne suis pas seul.\n"` |

Conclusion – a difficult font with little differentiation between "a" and "o", "l", "i" "L" and "t". a rather faded character "M" (in Mais) and an out of vocabulary word (OOV) – "gueltas" (a rock pool in the Hoggar) that doesn't even exist in the Grand Robert dictionary.

Google Cloud Vision makes the best attempt – the only mistakes are confusing lower-case "l" with upper-case. ABBY Finereader makes a creditable attempt.

This very brief sample confirms the results of other studies e.g. Tafti et al 2016 [22] that:

- Google Cloud Vision (and probably Microsoft Read and Amazon Textract/Rekognize) is best-in-class software-as-a-service
- Closely followed by the most reputed commercial software – ABBYY Finereader (and no doubt Kofax Omnipage)

These tools are considered to be several % points of accuracy better than open-source tools such as Tesseract. This is not surprising given the price-points of the dedicated tools or the resources used by the GA(FA)M. The authors have not had the time to assess all of the fine-tuning parameters of Tesseract to test this hypothesis.

For clean business documents these tools will give usable results – except for some special cases (such as some symbols or characters with unclear spacing).

For the rest of this small practical experiment it will be assumed that OCR issues in the real business world are insignificant and we will concentrate on the classification/extraction challenges using "real" PDF forms where possible.

## 8.3 Document classification

The next step in the IDP flow is the document classification challenge.

This is necessary in order to:

- Apply the most appropriate techniques for information extraction (IE)
- Divert the document/information to the right place in the workflow

This can either be a trivial problem or a very complicated problem.

To take the example of a supplier invoice if the only requirement is to identify that it is an invoice this can be done with a high degree of success by searching for a small number of keywords or combination of keywords (Invoice #; Facture du). But if the requirement is to identify the detailed class invoice/supplier the analysis may require more sophisticated techniques of Natural Language processing such as Named Entity Recognition.

The real-life rules can be quite complicated and require specific programming or Regex expressions. Bytescout PDF Multitools offers a simulation of the possibilities in the SDK – below in Fig. 5 is a simple case of Swiss administrative forms:

| Edit classes and their detection expressions | | | | |
|---|---|---|---|---|
| Class | Logic | Expression 1 | Expression 2 | Expression 3 |
| 3e pilier | OR | Form.21 EDP | prévoyance | |
| Certificat de salaire | OR | 605.040.18 Form 11 | salaire brut | |
| Facture | OR | Facture | | |
| | | | | |

**Fig. 5 – Simple key-word classification of Swiss administrative forms**

Classifying document images was for some time considered to be analogous to other computer vision tasks – i.e. a primarily a question of object recognition. Brants in 2003 [23] concluded that NLP, given the state-of-the-art at the time was not well suited to documental retrieval. Machine learning then boosted the progress in this field until around 2015-2016. However limitations from this approach were found due to frequent inter-class visual similarity or substantial intra-class heterogeneity.

Then more advanced techniques in NLP have come to the forefront – particularly the transformer model. The most recent developments combine CV and NLP in the same process (see the discussion in the State-of-the-Art section 10).

The same issues but amplified come up in Information Extraction so we will move on to that point with the specific example of supplier invoices

## 8.4 Information Extraction – special case of Supplier Invoices

Business forms share many similarities – they are structured or semi-structured (see previous definitions), zones of pure text are not extensive and the form has a "meaningful" structure of lines, columns, fields and tables.

This presents particular challenges for accurate information extraction.

A structured form – that has a fixed invariable format (such as the current year's tax form) does not present too many conceptual difficulties to extract the desired information. If the pdf form is digitally-born – whether it is a fillable form or a searchable downloadable copy of information entered into a web system the dimensions and field locations will be the same in all instances.

The desired fields can be identified by boundary boxes defined in x,y mm in a variety of software. The example below in Fig. 6 is a Swiss salary certificate with fictitious details and the boundary boxes defined manually in Bytescout PDF Multitool. The same technique of geometrically fixed bounding boxes can be used with OpenCV or GIMP or other tools.

The slightly more difficult case comes when a scanned version of this form has to be processed and the image is accordingly somewhat different in size and orientation compared to the original. If the image is taken by a smartphone camera it may even be deformed in a non-linear fashion.

Lohnausweis - Certificat de salaire - Certificato di salario
Rentenbescheinigung - Attestation de rentes - Attestazione delle rendite

808.21.972.943  220.3297.8872.26
AHV-Nr. - No AVS - N. AVS  Neue AHV-Nr. - Nouveau No AVS - Nuovo N. AVS

2003  30.8.2021  27.11.2013
Jahr - Année - Anno  von - du - dal  bis - au - al

Unentgeltliche Beförderung zwischen Wohn- und Arbeitsort
Transport gratuit entre le domicile et le lieu de travail
Transporto gratuito dal domicilio al luogo di lavoro

Kantinenverpflegung / Lunch-Checks
Repas à la cantine / chèques-repas
Pasti alla mensa / buoni pasto

7486627
Jamesina Mistry
Pilies Street
Wildomar, California

| | | Nur ganze Frankenbeträge<br>Que des montants entiers<br>Unicamente importi interi |
|---|---|---|
| 1. Lohn soweit nicht unter Ziffer 2-7 aufzuführen / Rente<br>Salaire qui ne concerne pas les chiffres 2 à 7 ci-dessous / Rente<br>Salario se non da indicare sotto cifre da 2 a 7 più sotto / Rendita | | 37853 |
| 2. Gehaltsnebenleistungen 2.1 Verpflegung, Unterkunft - Pension, logement - Vitto, alloggio | * | |
| Prestations salariales accessoires<br>Prestazioni accessorie al salario 2.2 Privatanteil Geschäfts--agen - P.privée voit. Serv. - Quota privata automobile di servizio | * | |
| 2.3 Andere - Autres - Altre<br>Art - Genre - Genere | * | |
| 3. Unregelmässige Leistungen - Prestations non périodiques - Prestazioni aperiodiche  Art - Genre - Genere | | |
| 4. Kapitalleistungen - Prestations en capital - Prestazioni in capitale | * | |
| 5. Beteiligungsrechte gemäss Beiblatt - Droits de participation selon annexe - Diritti di partecipazione secondo allegato | * | |
| 6. Verwaltungsratsentschädigungen - Ind. des membres de l'administration - Indennità dei membri di consigli d'amministrazione | * | |
| 7. Andere Leistungen - Autres prestations - Altre prestazioni<br>Art - Genre - Genere | * | |
| 8. Bruttolohn total / Rente - Salaire brut total / Rente - Salario lordo totale / Rendita | = | 59775 |
| 9. Beiträge AHV/IV/EO/ALV/NBUV - Cotisations AVS/AI/APG/AC/AANP - Contributi AVS/AI/IPG/AD/AINP | - | 1760 |
| 10. Berufliche Vorsorge 2. Säule 10.1 Ordentliche Beiträge - Cotisations ordinaires - Contributi ordinari<br>Prévoyance professionnelle 2° pilier<br>Previdenza professionale 2° pilastro 10.2 Beiträge für den Einkauf - Cotisations pour le rachat - Contributi per il riscatto | - | 2271 |
| 11. Nettolohn / Rente - Salaire net / Rente - Salario netto / Rendita<br>In die Steuererklärung übertragen - A reporter sur la déclaration d'impôt - Da riportare nella dichiarazione d'imposta | = | 53877 |
| 12. Quellensteuerabzug - Retenue de l'impôt à la source - Ritenuta d'imposta alla fonte | | -5957 |
| 13. Spesenvergütungen - Allocations pour frais - Indennità per spese<br>Nicht im Bruttolohn (gemäss Ziffer 8) enthalten - Non comprises dans le salaire brut (au chiffre 8) - Non comprese nel salario lordo (sotto cifra 8) | | |
| 13.1 Effektive Spesen 13.1.1 Reise, Verpflegung, Übernachtung - Voyage, repas, nuitées - Viaggio, vitto, alloggio<br>Frais effectifs<br>Spese effettive 13.1.2 Übrige - Autres - Altre<br>Art - Genre - Genere | | |
| 13.2 Pauschalspesen 13.2.1 Repräsentation - Représentation - Rappresentanza<br>Frais forfaitaires<br>Spese forfettarie 13.2.2 Auto - Voiture - Auto | | |
| 13.2.3 Übrige - Autres - Altre<br>Art - Genre - Genere | | |
| 13.3 Beiträge an die Weiterbildung - Contributions au perfectionnement - Contributi per il perfezionamento | | |
| 14. Weitere Gehaltsnebenleistungen Art<br>Autres prestations salariales accessoires Genre<br>Altre prestazioni accessorie al salario Genere | | |
| 15. Bemerkungen voir feuille additionnelle<br>Observations<br>Osservazioni | | |

I Ort und Datum - Lieu et date - Luogo e data
Genève, 21.6.2017

Die Richtigkeit und Vollständigkeit bestätigt
inkl. genauer Anschrift und Telefonnummer des Arbeitgebers
Certifie exact et complet
y c. adresse et numéro de téléphone exacts de l'employeur
Certificato esatto e completo
compreso indirizzo e numero di telefono esatti del datore di lavoro

Wessex Water
Louisburg Square
Granite Shoals, Texas

605.040.18 Form 11 (25.08.2008)

Seite - Page - Pagina 1 / 2

**Fig. 6 – a Swiss salary certificate with key fields to be extracted defined by bounding boxes**

Adrian Rosebrock et al of Pyimagesearch in their book OCR with OpenCV, Tesseract and Python – a Practitioner's Bundle (Chapter 6) [24] explain the technique and provide scripts to realign the image with the original document. In short this consists of identifying significant features on the master document and the derived document and then re-aligning the derived version to the master using a homography matrix and applying a perspective warp.

Then the fixed template can be successfully deployed as with the master document.

This is valid as long as the issuing organization does not change the format of the document – for example in the following tax year.

The same technique can be applied to specific families of invoices as long as the format remains identical across the family.

But how to adapt templates to forms with slightly different layouts?

Rosebrock in Chapter 7 of the book mentioned "we aren't creating a smart OCR system in which all text is recognized based on regular expression patterns. That is certainly doable but to keep this chapter lightweight I've manually defined….".

This statement resembles that of Pierre de Fermat in 1637 when he stated in a copy of Arithmetica that he had a proof of his own conjecture but that it was too large to fit in the margin. The theorem was finally proven in 2016. We can hope that the search for a smart form template does not take as long.

The specific case of supplier invoices has attracted a lot of research interest since the concrete business case is very obvious and the theoretical insights can be generalized to other difficult instances.

The great majority of invoices have the same 8-10 general concepts despite the apparent complexity of the 59 concepts shown on the sample Faktur-X invoice. The receiver will not need to identify his own name, address and identity details. Most invoices will not mention prepayments and most countries will not allow payments through varying forms of complicated promissory notes. So the key value types that need to be searched for can in most cases be reduced to a certain number of finite tables or codelists. Invoices do not contain much prose so the problem statement rather concerns extracting accurately a limited number of keys and their respective values. Nonetheless concepts such as payment methods can be very country-specific, e.g. postgiro or bankgiro in Sweden.

The difficulties arise from several sources:

- The commonly found information can be located at different places on every different supplier's invoice despite the attempts at standardisation
- The vendor's name and details can be found in different places on these documents even if there is a general preference to place this information at the top left of the form. Some suppliers place the client's details in this location and in general the right information for this value can get lost in the mass of information on the front page
- Extracting the line values from the invoice is even more challenging than the general information. These details are normally presented in tables and the challenge of extracting tables from documents has earned its own competition ICDAR2013 (see the next section 9) and leaderboard. Well-formed tables (perhaps 90%) can be solved but often tables may contain subsidiary nested tables, sub-headers or simply insufficiently clear distinctions between columns or lines. Needless to say the general problem is not yet completely solved – according to the F1 measure (see section 9 also) the best performing tools can achieve 90%-95% fulfilment at best.

It is possible to progress quite a long way on the general text information with rules and regular expressions, but tables are more difficult.

One example using Bytescout could be as shown in Table 4 below:

**Table 4 – some key fields to be extracted from a fictitious invoice**

| | |
|---|---|
| Invoice # | 00999 |
| Invoice Date | 12/12/2021 |
| Name of Rep. | Joe |
| Contactphone | 901-902-903 |
| Payment Terms | 30 days Net |

The environment (or culture) is defined as us-en and the date in American format MMDDYYYY.
The following Regular Expressions/macros can be used:

| | |
|---|---|
| InvoiceNumber | Invoice #\s+({{AnythingGreedy}}) |
| InvoiceDate | Invoice Date\s+((0[1-9]\|1[0-2]).(0[1-9]\|[12]\d\|3[01]).([12]\d{3})) |
| Name of Rep. | Name of Rep.\s+(\w+) |
| Contact Phone | Contactphone\s+({{LettersOrDigitsOrSymbols}}) |
| Payment Terms | Payment Terms\s+({{SentenceWithSingleSpaces}}) |

And the sample JSON below can be used for further processing as value pairs:

```json
{
    "templateName": "Test Company",
    "templateVersion": "4",
    "timestamp": "2021-12-25T11:43:36",
    "elapsed": 0.0383255,
    "objects": [
        {
            "name": "InvoiceNumber",
            "objectType": "field",
            "value": "00999",
            "pageIndex": 0,
            "rectangle": [
                0.0,
                0.0,
                0.0,
                0.0
            ]
        },
        {
            "name": "InvoiceDate",
            "objectType": "field",
            "value": "2021-12-12T00:00:00",
            "pageIndex": 0,
            "rectangle": [
                0.0,
                0.0,
                0.0,
                0.0
            ]
        },
        {
            "name": "Name of Rep.",
            "objectType": "field",
            "value": "Joe",
            "pageIndex": 0,
            "rectangle": [
                0.0,
                0.0,
                0.0,
                0.0
            ]
        },
        {
            "name": "Contact Phone",
```

Keys found below or above the relevant value need a bit more work but are solvable, e.g. the following fields in a French electricity invoice Fig. 7:

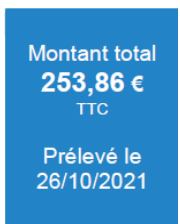

**Fig. 7 – 2 key fields to be extracted from a French electricity invoice**

Identifying vendor information may be trickier if it is not found in the common top-left position and may require a Named Entity Recognition approach.

If all of the required information is successfully retrieved and the remaining issue is to assign values to each key a scoring system may be used – for example the Hungarian algorithm which solves the assignment problem when X agents need to be assigned to Y tasks so that each task has one agent assigned and no agent is assigned to more than one task.

If a multi-country approach is needed some currency formats may also be difficult and require building a complicated set of rules. For example in Switzerland the decimal separator may be "." or "," and the thousands separator "'" or " ". There are also differences from one canton to another and between cantons and the federal level. The currency may be indicated as "CHF", "fr." (French), "Fr. (German) according to the usage and the language area and may precede or follow the currency value.

The authors will not therefore try and solve the problem here but simply review some significant work that has been done.

## 8.5 Tables

Identifying and correctly extracting tables is essential to the processing of many forms including invoices. Some significant publications on this subject are listed below:

**Nurimen (2013)** [25] already tackled this problem with an algorithmic approach seeking out proto-links between columns and rows. He treated only natively-digital PDFs using Poppler PDF API (NB LibreOffice also uses Poppler for PDF recognition and extraction). He used subsets of the ICDAR2013 dataset (EU and US) and achieved table structure recognition F1 scores of 96.57% and 86.85% respectively. The main issue with the algorithm was edge detection. The methodology did not deal well with elements at the top and bottom extremities and he did not tackle tables split over multiple pages. His algorithm has been used as a basis for the Tabula [26] software.

**Schreiber et al. (2017)** [27] demonstrated the DeepDeSRT method – Deep Learning approach for Table Recognition using a CNN network on both PDFs and images. The approach on ICDAR2013 yielded 96.77% for table detection and 91.44% for structure recognition. Better performance was indicated on PDFs compared to scanned images. On a closed dataset of a European aviation company 91.37% was achieved for recognition.

**Qasim et al. (2019)** [28] illustrated a methodology using graph neural networks and also showcased a new large-scale synthetic dataset. They tackled 4 different classes of tables with different levels of difficulty. The method seemed to outperform some other baselines but it is difficult to make a firm conclusion due to lack of comparability with some other approaches.

**Paliwal et al. (2019)** [29] with TableNet proposed a DL approach on scanned images with fine-tuning of the filters for rows and columns as well as an additional semantic layer. The results on ICDAR2013 gave 96.6% for table detection and 91.5% for table structure recognition.

**Pasad et al. (2020)** [30] demonstrate Cascade Tabnet – a CNN model with iterative transfer learning. As well as ICDAR datasets they have also used Marmot and Table Bank datasets. The results are difficult to interpret because of the caveats but seem to be in the low to middle 90s as a % F1 score.

## 8.6 Invoices – different approaches

Similarly a number of authors below have studied specifically invoices (which includes the problem of tables).

**Palm et al. (2017)** [31] proposed Cloudscan using RNNs. This is a single global model of invoice analysis that can be generalized to unseen invoices. For this they trained their system on a dataset of 33000 invoices and achieved 89.1% F1 scores on seen invoice layouts and 84% on unseen invoice layouts. Their system extracts automatically further training data from end-user provided feedback. They did not tackle line-items. Their approach provided superior results compared to a logistic regression classifier and an LSTM model. The trained embeddings were language agnostic. The main sources of error are OCR errors and discrepancies between the UBL (Universal Business Language) invoice and the PDF. Cloudscan has now been incorporated into the offering from Basware [32] (the Purchase-to-Pay specialists).

**Mayumder et al. (2020)** [33] – a team from Google Research – ingested both native PDF and scanned images through a cloud OCR service (Google Cloud Vision) to extract all of the text and then arrange it in a hierarchy. They then applied a schema using a cloud-based extraction service (Google Cloud NLP) and used a scoring and assignment model (that can be described as representation learning) to extract the correct candidate key for each value from a dataset of 14000 single-page invoices. At the field level the recall results varied from 87% to 99% and the precision results from 80% to 95%. They looked at not only invoices but also receipts which pose different challenges. Their approach was considered to be a challenging task since they looked at unseen templates.

**Holeček (2021)** [34] built a dataset with more than 25000 invoices and debit notes and applied a layout and proximity paring approach with a query-answer architecture. This was done on a consumer-grade GPU with only a few days training with subsequent fine-tuning and transfer-learning. The F1 score on key fields of this dataset was 92.9%. "Order ID" and "Terms" scored worst. Following his PhD Holeček went on to work for Rossum where he summarised "the teams are working hard but universal table extraction from invoices is still an unsolved problem".

**Patel et al. (2021)** [35] started with OCR for electronic invoices using AWS (Amazon Textract) that was subsequently exported into JSON. Textract creates the bounding boxes and useful information was extracted using Regex to then be sent back to the server to automatically fill up the PDF forms. Pre-processing was considered to be critical in this process. A confidence score of 98% was attributed to the extraction process – no assessment was made of the overall result including Regex. Future work includes automating the pre-processing step and allowing invoice captures from smartphones with a mobile app.

**León (2021)** [36] collated 900 crowdsourced Swedish invoices that were treated using Deep Learning techniques. Based on 4 critical fields his approach using NLP gave a F1 score of 91.1% compared to a rule-based method of 83.0% and Object Detection (or CV) of 81.5%.

Potential results of automatic processing of invoices should also mention the probable benefits from the SOTA techniques of treating documents with DL – see Section 10 for a short description of the latest developments. These leading-edge techniques are generally multimodal (combining NLP, CV and some rules) and require vast databases for pre-training as well as enormous computing power.

## 8.7 Conclusions for IDP of Invoices

From the available information (Web, publications) it can be concluded that streamlining the treatment of supplier invoices is a "hot" subject that not only involves a lot of current research but it likely to benefit from Computer Science developments generally and in particular machine learning.

Based on the apparent F1 scores for treating scanned or searchable-PDF forms some degree of confidence or automation becomes possible from around 85%-90% for relevant and accurate data extraction. State-of the-art techniques with some very specific fine-tuning can probably achieve 95% or more. However this still leaves some room for customization, human-in-the-loop and real-time training on errors found. Nonetheless one must not forget that humans are not error-free either – typically the annotated datasets used for training or verification contain 1%-2% of errors!!

## 9. Measures, training datasets and hardware environment

Given that the path to excellence in Computer Science is a kind of race and that its practitioners like to assess their progress it is understandable that a number of standard measures have evolved.

The most common for document classification and field extraction are Accuracy, Precision, Recall and F1 score (where F1 is the harmonic mean of Precision and Recall):

If a retrieved field, document or image is relevant it is considered as a True Positive (TP). Otherwise it is considered as a False Positive (FP).

If a relevant item is not retrieved it is counted as a False Negative (FN)

If an item is not relevant and it is not retrieved it is counted as a True Negative (TN)

$$Accuracy = \frac{TP+TN}{TP+TN+FP+FN}$$

$$Precision = \frac{TP}{TP+FP}$$

$$Recall = \frac{TP}{TP+FN}$$

$$F1\ Score = \frac{2*Precision*Recall}{Precision+Recall}$$

Machine learning thrives on data (otherwise it is just computer programming!) so naturally a number of datasets have been developed for training and validation. Many of these are annotated by humans as "ground truth". Some are publicly available datasets, some are private or confidential.

The largest public database – in the English-speaking world at least - is the IIT-CDIP (Illinois Institute of Technology – Complex Document Information Processing). This was established and annotated as part of the lawsuits against the American Tobacco industry. In total there are 11 million documents in 16 categories that have been listed previously in this paper. Subsets of this same dataset are RVL (Ryerson Vision Lab)-CDIP also called "Big Tobacco" and a smaller sub-subset called Small Tobacco. This data set is often used for pre-training before fine-tuning on more specific domain datasets.

Other datasets are used for competition comparison purposes, for example:
- ICDAR2013 for table detection and extraction (ICDAR = International Conference on Document Analysis and Recognition)
- GLUE benchmark (General Language Understanding Evaluation) for NLP

Some other datasets are private for confidential or competitive reasons.

For wordsets the best-known is WordNet which includes much of the corpus of the English language with synonyms and relationships and has free public access. For European languages an attempt was made to create a similar glossary/thesaurus for all of the European languages – EuroWordNet – but development was stopped in 2000 and access requires payment of a license. Europe is not ready to win the race in Computer Science with such an approach!

For hardware the arms-race is also underway especially in machine-learning. Training especially has moved on from consumer-grade CPUs (Central Processing Units) to GPUs (Graphical Processing Units) often supplied by NVIDIA. The authors have made a few comparative tests on some simple processes of ML and reached results of 60x on a GPU compared to a CPU. The next step was multiple ranks of GPUs and now some processes are run on TPUs (Tensile Processing Units) – specific chipsets developed for Deep Learning with a power of several petaflops (10 to the power of 15 floating point operations per second). The benefits of this deployment by the big multinationals and research institutes is that the results trickle down to the rest of the community and are often made public.

Google Colaboratory [37] for example allows researchers to write and execute Python in a browser with zero configuration and free access to GPUs.

## 10. Historic development and State-of the-Art

Herein follows a short summary of some of the methods that have been deployed in IDP – especially given the staggering recent development in AI.

Up to 2000 the prevalent mode in industry was undoubtedly rules, heuristics and hand-crafted code. In practice this may still be the case in many firms.

From around 2000, in theoretical research at least, machine learning and then deep-learning attracted the majority of interest.

Earlier techniques were mainly algorithm-based:

- Logistic or Linear Regression (statistical) [38]
- Feature Detectors such as SIFT (Scale-Invariant Feature Transform) [39] and SURF (Speeded-Up Robust Features) [40]
- Decision Trees and Random Forest (ensembles of independent trees) [41]
- Naïve Bayes (Probabilistic) [42]
- K-Nearest Neighbours (statistical/clustering) [43]
- Support Vector Machine (High Dimension Classification and Regression) [44]
- Hidden-Tree Markov Model  (Probabilistic) [45]

The development of neural techniques brought about new configurations and techniques:

- RNN (Recurrent Neural Network) where nodes form a graph along a temporal sequence where the network has an infinite impulse response through a directed cyclic graph that cannot be unrolled
- LSTM (Long Short-Term memory) is a type of RNN that processes not only single data points but sequences of data.
- CNN (Convolutional Neural Network) where the graphs have a finite impulse response through a directed acyclic graph that cannot be unrolled

These concepts become difficult to understand in layman terms. Convolutional = mathematical combination of two functions to produce a third function.

In practice RNNs are often largely used in text analysis and CNNs in visual imagery.

Up until around 2015 visual imagery or object detection seemed, at least in theoretical publications, to be the predominant technology for processing forms.  The limitations began to be seen of applying only visual recognition as explained previously. Thereafter NLP began to re-emerge with new technologies:

- The Transformer model which adopts the mechanism of self-attention which gives different weight to different parts of the input data [46] [47]
- BERT (bi-directional encoder representations from transformers) [48] developed by Google which applies bidirectional training to language modelling which therefore better takes into account the context

BERT gave rise to further developments such as RoBERTa etc.

As a natural result of these different streams consideration began to be given to multimodal deep networks that combine both CV and NLP approaches (**Audebert et al 2019** [49] or **Bakhali et al 2020** [50]).

The current SOTA seems to be recent developments by Microsoft research that combines an analysis of the layout of the form or its visual aspect with a semantic or NLP approach:

LayoutLM (2020) [51] or

LayoutXLM (2021) [52] – a multilingual version

All of these neural models are pre-trained and then fine-tuned.

The concept of training and then deployment in a different domain is called transfer-learning.

The pre-training is carried out on semi-relevant annotated data or even unannotated data. If the latter idea is counter-intuitive it can be assimilated to making a "brain" agile or receptive [53].

The real training or fine-tuning should be carried out on an annotated dataset – the most relevant possible to the real use-case.

Testing is carried out on an unseen dataset and validation against another annotated dataset.

Iterative training can be injected into the training process.

In addition, to take multimodal a step further an additional layer of semantic richness can be added such as an ontology. Fastext (developed by Facebook) is sometimes mentioned as an additional tool (not investigated).

There are several concerns about Machine Learning that may be raised by those who are not involved in this field:

1. How to secure and then transfer the benefits of the learning and tuning of the model. It would seem that saving and then reloading the acquired model is done through serialisation and deserialisation. Serialisation saves the model including all the parameters and weighting in a number of files including a (potentially huge) hdf5 file. Deserialising reverses that process. Deploying into another environment can be done through "containers" that encapsulate both the model and the environment necessary to run it. These can be connected to other applications via a REST (representational state transfer) API that is environment and programming language agnostic. Google (Kubernetes), Docker (Swarms) and Amazon (Elastic Services all have offers in the domain of containerization.

2. Understanding and interpreting what the machine is actually doing. This may seem possible with simple algorithms but Machine Learning rapidly becomes a "black box".  Google and Amazon (and no doubt other firms) have products that carry out audit trails or a step-by-step analysis of the learning progress.

3. The fact that, even if pre-training can be carried out on other semi-relevant databases, the live training and  fine-tuning needs to be carried out on a significant relevant database. For example in France if one wants to automate extraction of documents received from complementary medical insurers one needs to take into account that there are 435 different organisations. So a CV approach will not work based on one or two samples and unless an NLP or Bag-of-Words approach can cover all of the cases specific templates seem to be difficult to develop.  **Esser et al. (2014)** [54] with "Few-Exemplar Information Extraction for Business Documents" argue differently and maintain that it is possible to achieve reasonable results with initially few documents and a rapid learning curve but this hypothesis has not been extensively tested in the literature.

## 11 The Commercial market for IDP

It is difficult to identify the commercial market for IDP since it is evolving constantly and may be mixed up with different or broader concepts such as Document Management Software or RPA.

Nor is it easy to separate the marketing hype from reality or the "froth" from the beer.

According to marketsandmarkets [55] the Intelligent Document Processing Market is expected to grow from $B 0.8 in 2021 to $B 3.7 in 2026.

Gartner [56] assesses the market as $B 1.2 in 2020.

Another approach is to look at the size of some of the players in this and overlapping markets.

- Uipath has a turnover of $M 727
- Kofax has a turnover of approx. $M 500
- ABBYY has a turnover of approx. $M 150

These companies all have wider scopes than purely IDP.

The smaller companies and start-ups often have turnover of the order of single digit $M. Rossum has a turnover of around $M 10.

The turnover figures are indicative of the resources that these companies can deploy for a specific implementation project.

A further approach is to look at the offers of SaaS proposed by the big multinational companies (GAM…) – they all have highly developed recent offers scalable into the millions of documents per year which indicates that the market need for IDP for large corporations exists and is indeed used.

In summary it is very easy to believe that the market is well in excess of $1B and increasingly rapidly.

The countervailing force is the extent to which paper-based or "dumb" pdf documents will be replaced completely and will be replaced by automatically interchangeable formats and information.

The main brake on the uptake by larger companies is undoubtedly the weight of legacy systems, and the cost of adaptation and integration.

## 12 Conclusion

It is not easy to summarise this long and winding road through the landscape surrounding Intelligent Document Processing. The challenge for industry and researchers is not new.

Already in 2013 Chiticariu et al. (of IBM Research) [57] concluded "we surveyed the landscape of IE techniques and identified a major disconnect between industry and academia: while rule-based IE dominates the commercial world it is widely regarded as dead-end technology by the academia".

Has the world changed since 2013? In terms of readiness the authors believe so, but practically?

The somewhat personal takeaways from this survey are as follows:

- The concerns of practitioners in the real world are somewhat different from those of Computer Science researchers. Industrialists are looking for a concrete result with defined budgets and time-lines. Researchers, generally speaking, are pursuing progress in less-applied and more long-term knowledge-domains. At this moment the most impressive progress in fundamental techniques seems to be accomplished by the GAFM
- In the real world most automated solutions for classifying, extracting or forward-processing documents use to a large extent commercial products or hybrid solutions. One can find a cohabitation of Computer Vision, Natural Language Processing, rules-based approaches or handcrafted programs along with human intervention. Researchers are, for the most part, focused on Open Source tools.
- The concerns of a multinational company to integrate millions of documents into its ERP system with no tolerance for errors or a private individual or SME who is just trying to improve its ability to better classify or find a small number of documents are very different
- Machine Learning approaches require vast quantities of data and documents in order to achieve good results. Training on publicly-available datasets is often of limited value compared to documents originating from the real use-case or domain of interest

The authors hope that this (relatively) brief walk-through this subject will enable:

- Industrialists to have a better understanding of the language and the methods of Computer Scientists
- Computer Scientists to have a better understanding of the perspective and needs of industrialists (and private individuals)

## 13 Future work

Much remains to be done.

The authors' personal wish-list includes:

- The "Fermat-like" problem of producing a smart template for document information extraction
- Seeing the problem of table extraction solved – the SOTA tools are beginning to come close
- Producing a Decision-Support System in order to guide decision-makers in building a use-case and identifying the most promising technologies or suppliers
- Carry out an in-depth case study of a real implementation project of IDP in a large industrial company. This sort of a study is often carried out in business schools and, without revealing all the trade secrets it should be possible to gain insights into the degree of customization and the implementation effort involved
- Perform a more detailed review of the customer stories available on Internet (there are several hundred available). Although they are normally brief (2-3 pages) a detailed reading can often reveal the business sector, the process and some of the techniques used. This may not be statistically total representative but would give a better idea of the profile of applications in the commercial world
- Seeing a better and free glossary/thesaurus developed for the main European languages
- Open access to a free Regex library that identifies all of the possible date and currency variants

# 14 Bibliography and references

## Biography of authors

### Graham Cutting

M.A. in Economics from Cambridge University (U.K.)

Various financial management positions with Massey-Ferguson in Coventry (U.K.) (1976-1988) including Factory Chief Accountant with Purchase-to-Pay responsibilities

Finance Manager for Massey-Ferguson's European Spare Parts Operation (1988-1992) Athis-Mons (France) with operational locations also in Barcelona (Spain) and Copenhagen (Denmark).

Finance Administration and IT Manager for part of SKF (world leader in ball and roller bearings) – specifically for Transrol S.A.S. and Linear Motion group (1988-2003). Chambéry (France) and Torino (Italy). Involvement in financial appraisal of several acquisitions.

Division Business Controller for SKF (2003-2014) Brussels (Belgium) and Gothenburg (Sweden) with follow-up/audit responsibilities for more than 100 subsidiaries and representative on corporate Process Board. Involvement, in several ERP projects (Movex, SAP) including Project Leader responsibilities.

Several consulting projects carried out subsequently including product assortment/metadata and acquisition appraisal.

Private research in establishing a state-of-the art worldwide referenced and geolocalised database of cartography – both historical and modern. This often involves converting file formats, installing or correcting Coordinate Reference Systems or translating to/from many languages such as Japanese, Arabic…

**Anne-Françoise Cutting-Decelle**

Prof. Anne-Francoise Cutting-Decelle obtained her PhD at the Civil Engineering and Habitat Laboratory of the University of Savoie Mont-Blanc, Chambery, France. She was previously Professor in Software engineering at the Ecole Centrale de Lille, she is now a researcher at the Computer Science Center of the University of Geneva, Switzerland. Her research area focuses on the adaptation of knowledge engineering and information extraction techniques to improve understanding and interoperability of international standards used in industry, through the use of ontologies and natural language processing techniques, with the objective to facilitate the transfer from human readable information to machine processable data